\documentclass[12pt]{iopart}
\usepackage{iopams}
\usepackage{setstack}
\usepackage{psfig}
\newcommand{\half}{\mbox{$\textstyle \frac{1}{2}$}}

\newcommand{\re}{{\rm e}}
\newcommand{\ri}{{\rm i}}
\newcommand{\rd}{{\rm d}}
\newcommand{\unit}{\mkern1mu \lower0.2pt \hbox{1} \! \! 1}

\newcommand{\bard}[1]{\bar{#1}}

\begin{document}
\title[Microcanonical distributions for quantum
systems]{Microcanonical distributions for quantum systems}

\author[Brody, Hook, and Hughston]{Dorje~C~Brody$^{*}$,
Daniel~W~Hook$^{*}$, and Lane~P~Hughston$^{\dagger}$}

\address{${}^{*}$Blackett Laboratory, Imperial College,
London SW7 2BZ, UK}

\address{${}^{\dagger}$Department of Mathematics, King's College
London, London WC2R 2LS, UK}

\begin{abstract}
The standard assumption for the equilibrium microcanonical state
in quantum mechanics, that the system must be in one of the energy
eigenstates, is weakened so as to allow superpositions of states.
The weakened form of the microcanonical postulate thus asserts
that all quantum states giving rise to the same energy expectation
value must be realised with equal probability. The consequences
that follow from this assertion are investigated. In particular, a
closed-form expression for the density of states associated with
any system having a nondegenerate energy spectrum is obtained. The
result is applied to a variety of examples, for which the
behaviour of the state density, as well as the relation between
energy and temperature, are determined. Numerical studies indicate
that the density of states converges to a distribution when the
number of energy levels approaches infinity. (\today)
\end{abstract}

\submitto{\JPA}

%
%

\section{Introduction}
\label{sec:1}

It is sometimes argued that the quantum-mechanical description of
the microcanonical distribution is simpler than the corresponding
classical counterpart~\cite{thompson}. This is because the
standard quantum microcanonical postulate asserts that if the
energy of the system lies in the range $E$ to $E+\Delta E$, then
all of the energy eigenstates with energy $E_n\in[E,E+\Delta E]$
are realised with equal probability~\cite{huang}. While such a
postulate does indeed provide an elementary statistical
description of the quantum system in terms of a uniform
probability distribution, it is not clear how this postulate ties
in with other fundamental ideas in quantum mechanics.

For example, in quantum mechanics the state of the system can be
represented by a general superposition of energy eigenstates,
whereas according to the standard quantum microcanonical
postulate, superpositions of energy eigenstates with distinct
eigenvalues are excluded. Also, if an energy eigenvalue that lies
in the interval $[E,E+\Delta E]$ is nondegenerate, then the
standard postulate implies that the system can only be in a single
eigenstate corresponding to that eigenvalue. These observations
suggest that the standard postulate is perhaps too stringent to
give rise to a satisfactory statistical description of a quantum
system in isolation.

The purpose of the present paper is to study the consequences of a
relaxation of the standard quantum microcanonical postulate.
Specifically, we consider the following generalisation of the
microcanonical postulate: namely, that every quantum state
possessing the same energy expectation value must be realised with
an equal probability. According to this `weaker' postulate various
superpositions of energy eigenstates with distinct energies are no
longer excluded; likewise, a microcanonical state will never
correspond to a single eigenstate (except in the case of the
smallest and the largest energy levels if these are
nondegenerate).

The idea that we investigate here is similar to the classical
case, although there is a subtle difference. Classically, the
uncertainty in energy is fully characterised by a statistical
distribution over the phase space, and for a classical
microcanonical distribution having support on a level surface of
the Hamiltonian the energy variance vanishes. Quantum
mechanically, the contribution to the energy variance from the
statistical distribution over the phase space also vanishes.
However, if the specified energy level is not the largest or
smallest energy eigenvalue, then there is an additional
contribution to the energy variance that arises from the intrinsic
quantum uncertainty. Therefore, according to our quantum
microcanonical postulate the energy variance of a system described
by a quantum microcanonical ensemble need not vanish, even though
the system is in isolation.

The paper is organised as follows. In \S\ref{sec:2} we review
briefly the formulation of standard quantum theory in terms of the
geometry of the space of pure states, as described, e.g., in
\cite{anandan,ashtekar,brody,cirelli,hughston,kibble} and
references cited therein. When the trajectory of a wave function
is projected from Hilbert space ${\mathcal H}$ to the space
${\mathcal P}$ of pure states (rays through the origin of
${\mathcal H}$), the Schr\"odinger equation on ${\mathcal H}$
reduces to Hamilton's equation on ${\mathcal P}$. More precisely,
the space of pure states ${\mathcal P}$ has a natural symplectic
structure; and the Schr\"odinger trajectories, when projected onto
${\mathcal P}$, are the integral curves of the Hamiltonian vector
field obtained by taking the symplectic gradient of the function
on ${\mathcal P}$ defined by the expectation of the Hamiltonian
operator. As a consequence, the formulation of quantum mechanics
on ${\mathcal P}$ provides a natural environment in which one can
study important issues arising in the context of Hamiltonian
mechanics, such as ergodicity conditions or the construction of
equilibrium ensembles. For example, the dynamical approach to
microcanonical equilibrium introduced by Rugh~\cite{rugh} for
classical systems can be seen to apply in the quantum regime.

In \S\ref{sec:3} we introduce the microcanonical density of states
that follows from our weakened microcanonical postulate, together
with the associated microcanonical density matrix. In
\S\ref{sec:4} we derive a general integral representation for the
density of states, expressed in terms of the energy eigenvalues.
We then perform the integration explicitly in \S\ref{sec:5} in the
case for which the energy spectrum is nondegenerate. As an
illustration, the properties of a system having an equally spaced
energy spectrum are studied in detail in \S\ref{sec:6}. In this
example we study the relation between the energy and the
temperature in some detail, which we plot for a number of
different situations. In particular, a simple procedure for
rescaling the energy shows that as the dimensionality of the
Hilbert space increases the system becomes more and more likely to
take the intermediate energy value $\frac{1}{2}(E_{\rm max}+E_{\rm
min})$, where $E_{\rm max}$ and $E_{\rm min}$ are, respectively,
the largest and the smallest energy eigenvalues. To study the
convergence of the distribution numerically, we compute in
\S\ref{sec:7} the Hellinger distance between the density of states
associated with an $n$-level system and an $(n+1)$-level system
for $n=2,3,\ldots$. The result shows that the logarithmic plot of
the relative distances of the distributions against the Hilbert
space dimensionality lies on a straight line with gradient $-2$.
Finally in \S\ref{sec:8} we study the properties of systems having
other nondegenerate spectra. In particular, we show numerically
that there is an approximate symmetry relation that holds between
a system having the spectrum $E_n\sim n^k$ and a system having the
spectrum $E_n\sim n^{1/k}$, where $k$ is a constant.

\section{Quantum phase space}
\label{sec:2}

In order to investigate properties of a closed, isolated quantum
system in equilibrium it will be useful first to recall the
Hamiltonian formulation of standard quantum mechanics. This has
the advantage of allowing us to apply concepts arising in the
corresponding classical theory of equilibria as outlined, for
example, in Ref.~\cite{ehrenfest}.

Physical states in quantum mechanics are represented by elements
of a complex Hilbert space ${\cal H}$, which we assume to be
$(n+1)$-dimensional. Let us denote by $Z^\alpha$ a typical element
of ${\cal H}$, so that the index $\alpha$ runs over the range
$\alpha=0,1,\ldots,n$. The Hamiltonian, which acts on elements of
${\cal H}$, can thus be denoted $H^\alpha_\beta$, and the
expectation value of $H^\alpha_\beta$ in the state $Z^\alpha$ is
given by the expression
\begin{eqnarray}
\langle H\rangle = \frac{{\bar Z}_\alpha H^\alpha_\beta
Z^\beta}{{\bar Z}_\gamma Z^\gamma}, \label{eq:1}
\end{eqnarray}
where ${\bar Z}_\alpha$ is the complex conjugate of $Z^\alpha$.
The Hilbert space ${\cal H}$ carries an essentially irrelevant
complex degree of freedom given by the overall scale of the state
vector. This follows from the fact that the expectation value of a
physical observable is invariant under the complex scale
transformation $Z^\alpha\to\lambda Z^\alpha$, where $\lambda\in
{\mathbb C}-\{0\}$. It is useful in some applications to eliminate
this extra degree of freedom by considering the space of
equivalence classes under the relation $\lambda Z^\alpha\sim
Z^\alpha$ for $\lambda\in {\mathbb C}-\{0\}$. This is the space of
rays through the origin of ${\cal H}$, otherwise known as the
projective Hilbert space ${\cal P}$ of complex dimension $n$. It
is well known that quantum theory, when formulated on the
projective space ${\mathcal P}$, admits a representation in terms
of the standard mathematical structure of Hamiltonian mechanics.
This can be seen as follows.

We find it convenient for our purposes to regard the projective
space ${\mathcal P}$ as a real manifold $\Gamma$ of dimension
$2n$, letting $x^a$ $(a=1,2,\ldots,2n)$ denote a typical point in
$\Gamma$. Therefore, each point $x^a\in\Gamma$ represents a ray in
the Hilbert space ${\mathcal H}$. In this way we can regard the
expectation (\ref{eq:1}) as determining a real-valued function
$H(x)$ on $\Gamma$.

The space of pure states, when regarded as the real
even-dimensional space $\Gamma$, is endowed with a symplectic
structure, given by a nondegenerate, antisymmetric two-form
$\omega_{ab}$. The dynamical laws governing the trajectories of
quantum states, given by the Schr\"odinger equation on ${\mathcal
H}$, can then be represented on $\Gamma$ in Hamiltonian form as
follows:
\begin{eqnarray}
\half\hbar\omega_{ab} \frac{\rd x^b}{\rd t}= \nabla_aH(x).
\label{eq:2}
\end{eqnarray}
In other words, the space $\Gamma$ is a symplectic manifold upon
which the evolution of a quantum state is governed by Hamilton's
equations, which in the language of symplectic geometry take the
form (\ref{eq:2}). Therefore, we can regard $\Gamma$ as the
quantum analogue of a classical phase space. As a consequence we
can also formulate our investigation of equilibrium states on
$\Gamma$.

\section{The microcanonical ensemble}
\label{sec:3}

We begin this section by considering the foliation of the quantum
phase space $\Gamma$ by level surfaces of the Hamiltonian function
$H(x)$. This is given by a family of hypersurfaces $\{{\cal
E}_{E}\}$, $E\in[E_{\rm min},E_{\rm max}]$, determined by level
values $H(x)=E$ of the Hamiltonian function. The structure of the
typical energy surface in quantum mechanics is quite intricate,
even for a system described by low-dimensional Hilbert space. In
particular, as $E$ varies in the given range $[E_{\rm min},E_{\rm
max}]$ both the dimensionality and the topology of the associated
energy surfaces can change. An example can be found in
Ref.~\cite{brody}, in which the structures of the energy surfaces
for a three-level system are investigated in detail.

Now given this foliation, the `number' of quantum mechanical
microscopic configurations (pure states) with expected energy in
the small range $E$ and $E+\Delta E$ is $\Omega(E)\Delta E$, where
the state density $\Omega(E)$ for energy $E$ is given by an
expression of the form
\begin{eqnarray}
\Omega(E) = \int_{{\cal E}_{E}} \frac{\nabla_{a}H \rd \sigma^{a}}
{\nabla_{b}H\nabla^{b}H}\ . \label{eq:3}
\end{eqnarray}
Here the natural vector-valued ($2n-1$)-form $\rd\sigma^a$ on
$\Gamma$ is defined by
\begin{eqnarray}
\rd\sigma^{a}=g^{ab}\epsilon_{bc\cdots d}\, \rd x^{c}\cdots \rd
x^{d} , \label{eq:4}
\end{eqnarray}
where $\epsilon_{bc\cdots d}$ denotes the totally skew tensor with
$n$ indices, and $g^{ab}$ is the inverse of the natural Riemannian
metric $g_{ab}$ on $\Gamma$. The metric $g_{ab}$ is compatible
with the symplectic structure $\omega_{ab}$ in the sense that
$\nabla_{a}\omega_{bc}=0$, where $\nabla_a$ is the unique
torsion-free covariant derivative operator on $\Gamma$ satisfying
$\nabla_ag_{bc}=0$. It is a remarkable feature of the quantum
phase space $\Gamma$ that it has both a natural Riemannian metric
and a compatible symplectic structure. These elements can be
regarded as part of the natural geometry of any quantum system.

In the case of an isolated quantum mechanical system with energy
in the small range $E$ to $E+\Delta E$, we can adopt the notion of
the microcanonical ensemble in classical statistical mechanics
(cf.~\cite{thompson}), and identify the entropy of the system by
use of the Boltzmann relation
\begin{eqnarray}
S(E) = k\ln(\Omega(E)\Delta E). \label{eq:5}
\end{eqnarray}
Here we are implicitly assuming what might be called the {\it
quantum microcanonical postulate}, which asserts that {\it for an
isolated system in equilibrium all states on a given energy
surface in the quantum phase space are equally probable}. As a
consequence, the temperature $T$ of such a system is determined by
the relation
\begin{eqnarray}
\beta = \frac{\rd S(E)}{\rd E}, \label{eq:6}
\end{eqnarray}
where $\beta=1/kT$ and $k$ is Boltzmann's constant. Thus for an
isolated quantum system with expected energy $E$, we assume that
the equilibrium configuration is given by a uniform distribution
on the energy surface ${\cal E}_{E}$, with entropy $S(E)$ and
inverse temperature $\beta(E)$, as given above. The corresponding
probability density on ${\sl\Gamma}$, which we call the
microcanonical $\Gamma$-distribution, is
\begin{eqnarray}
\mu_{E}(x)\ =\ \frac{1}{\Omega(E)}\,\delta(H(x)-E), \label{eq:7}
\end{eqnarray}
where
\begin{eqnarray}
\Omega(E)=\int_{\Gamma}\delta(H(x)-E)\rd V. \label{eq:8}
\end{eqnarray}
Here $\rd V$ is the volume element on $\Gamma$.

It is a straightforward exercise to show that starting from the
definition (\ref{eq:8}) of the density of states $\Omega(E)$ we
can deduce the integral formula (\ref{eq:3}). This can be seen as
follows (cf.~\cite{khintchin}).

First we note that at each point $x\in\Gamma$ such that
$\nabla_aH(x)\neq0$ the volume element on $\Gamma$ can be written
as a product
\begin{eqnarray}
\rd V=\rd N \rd\sigma,
\end{eqnarray}
where the $(n-1)$-form $\rd\sigma$ defined by
\begin{eqnarray}
\rd \sigma = \frac{\nabla^aH\epsilon_{ab\cdots c}\rd x^b\cdots \rd
x^c}{\sqrt{\nabla_dH\nabla^dH}}
\end{eqnarray}
is the volume element on the energy surface passing through $x$,
and
\begin{eqnarray}
\rd N = \frac{\nabla_aH\rd x^a}{\sqrt{\nabla_bH\nabla^bH}}.
\end{eqnarray}
On the other hand, as a consequence of the relation
\begin{eqnarray}
\rd H=\nabla_aH\rd x^a
\end{eqnarray}
we observe that
\begin{eqnarray}
\rd N = \frac{\rd H}{\sqrt{\nabla_aH\nabla^bH}}.
\end{eqnarray}
Substituting this expression into (\ref{eq:8}) we obtain
\begin{eqnarray}
\Omega(E)&=&\int_{\Gamma}\delta(H(x)-E)\rd V \nonumber \\ &=&
\int_{\Gamma}\delta(H(x)-E)\frac{\rd\sigma \rd H}
{\sqrt{\nabla_aH\nabla^bH}} \nonumber \\ &=& \int_{{\cal E}_{E}}
\frac{\rd \sigma}{\sqrt{\nabla_{b}H\nabla^{b}H}} \nonumber \\ &=&
\int_{{\cal E}_{E}} \frac{\nabla^aH\epsilon_{ab\cdots c}\rd
x^b\cdots \rd x^c}{\nabla_dH\nabla^dH} ,
\end{eqnarray}
which agrees with expression (\ref{eq:3}).

A general measurable function $F(x)$ on $\Gamma$ represents a
nonlinear observable in the sense of Kibble~\cite{kibble} and
Weinberg~\cite{weinberg}. The usual `linear' observable of
standard quantum mechanics then corresponds to the situation for
which $F(x)$ can be represented as the expectation of a Hermitian
operator. In either case, for each value of $x$ we interpret
$F(x)$ as the conditional expectation $\langle F\rangle_{x}$ of
the observable $F$ in the pure state $x$. The unconditional
expectation of $F$ in the microcanonical ${\Gamma}$-ensemble is
then given by
\begin{eqnarray}
\langle F\rangle_{E} = \int_{\Gamma}F(x)\mu_{E}(x)\rd V.
\label{eq:9}
\end{eqnarray}
In the case of a linear observable we have
\begin{eqnarray}
F(x) = F_{\alpha}^{\beta} \Pi_{\beta}^{\alpha}(x),
\end{eqnarray}
where
\begin{eqnarray}
\Pi_{\beta}^{\alpha}(x) = \frac{{\bar Z}_\beta Z^\alpha}{{\bar
Z}_\gamma Z^\gamma} \label{eq:10}
\end{eqnarray}
is the projection operator onto the state vector $Z^\alpha(x)$
corresponding to the pure state $x\in\Gamma$. Then the
unconditional expectation in the state $\mu_{E}(x)$ is
\begin{eqnarray}
\langle F\rangle_{E}=F_{\alpha}^{\beta} \mu^{\alpha}_{\beta}(E),
\end{eqnarray}
where the quantum microcanonical density matrix
$\mu^{\alpha}_{\beta}(E)$, parameterised by $E$, is defined by
\begin{eqnarray}
\mu^{\alpha}_{\beta}(E) = \int_{\Gamma}
\Pi^{\alpha}_{\beta}(x)\mu_{E}(x)\rd V . \label{eq:11}
\end{eqnarray}
Providing we only consider linear observables, i.e. as in standard
quantum mechanics, the state of the system is then fully
characterised by the density matrix $\mu^{\alpha}_{\beta}(E)$.

Now suppose $W(E)$ denotes the total phase space volume for states
such that $H(x)\leq E$. Then the density matrix
$\mu^{\alpha}_{\beta}(E)$ can be calculated explicitly by use of a
`variation-of-parameters' formula given by
\begin{eqnarray}
\mu^{\alpha}_{\beta}(E) =-\left(\frac{\rd W(E)}{\rd E}\right)^{-1}
\frac{\partial W(E)}{\partial H^{\beta}_{\alpha}}. \label{eq:12}
\end{eqnarray}
This representation can be verified as follows. From the
definition of $W(E)$ we can write
\begin{eqnarray}
W(E) = \int_{-\infty}^E \int_\Gamma \delta\left(H^\beta_\alpha
\Pi^\alpha_\beta (x)- u\right) \rd V \rd u,
\end{eqnarray}
and hence
\begin{eqnarray}
\frac{\partial W(E)}{\partial H^{\beta}_{\alpha}}&=&
\int_{-\infty}^E \int_\Gamma \Pi^\alpha_\beta(x) \delta^\prime
\left( H^\beta_\alpha \Pi^\alpha_\beta (x)- u\right) \rd V \rd u
\nonumber \\ &=& \int_\Gamma \Pi^\alpha_\beta(x)\left(
\int_{-\infty}^E \delta^\prime\left( H^\beta_\alpha
\Pi^\alpha_\beta (x)- u\right) \rd u\right) \rd V \nonumber \\ &=&
- \int_\Gamma \Pi^\alpha_\beta(x) \,\delta\left(
H^\beta_\alpha \Pi^\alpha_\beta (x)- E\right) \rd V \nonumber \\
&=&-\Omega(E)\int_{\Gamma}\Pi^{\alpha}_{\beta}(x)\mu_{E}(x)\rd V
\nonumber \\ &=& -\Omega(E) \mu^\alpha_\beta(E).
\end{eqnarray}
On the other hand, clearly $\rd W(E)/\rd E=\Omega(E)$, and thus we
obtain (\ref{eq:12}).

\section{Calculating the density of states}
\label{sec:4}

As defined in equation (\ref{eq:8}) above, the density of states
is given by the volume integral over $\Gamma$ of a delta-function
having a support on the energy surface ${\mathcal E}_E$. Our
objective now is to perform the relevant integration explicitly
for a generic Hamiltonian, and obtain a representation for
$\Omega(E)$ in terms of the energy eigenvalues.

We find it convenient to pursue the calculation by lifting the
integration from the phase space $\Gamma$ to the Hilbert space
${\mathcal H}$, imposing the constraint that the norm of the
Hilbert space vector $Z^\alpha$ is unity. Therefore, we write the
expression (\ref{eq:8}) in the following form:
\begin{eqnarray}\label{eq:int2}
\Omega(E)=\frac{1}{\pi}\int_{{\mathbb C}^{n+1}} \delta( {\bar
Z}_\alpha Z^\alpha-1)\;\delta\left(\frac{{\bar Z}_\alpha
H^\alpha_\beta Z^\beta}{{\bar Z}_\gamma Z^\gamma}-E\right) {\rm
d}^{n+1}\bard{Z}\; {\rm d}^{n+1}Z. \label{eq:13}
\end{eqnarray}
The additional factor of $\pi$ arises here from the superfluous
phase integration in (\ref{eq:13}). By use twice of the standard
integral representation
\begin{eqnarray}
\delta(x)=\frac{1}{2\pi} \int^{\infty}_{-\infty} \re^{-\ri\lambda
x} \rd\lambda  \label{eq:14}
\end{eqnarray}
we thus deduce that
\begin{eqnarray}
\Omega(E)&=&\frac{1}{\pi}\int^{\infty}_{-\infty} \frac{{\rm d}
\lambda}{2\pi} \int^{\infty}_{-\infty} \frac{{\rm d}\nu}{2\pi}
{\rm e}^{{\rm i}(\lambda+\nu E)} \nonumber \\ & & \times
\int_{{\mathbb C}^{n+1}} \exp\Big[(-{\rm i}\left(\lambda {\bar
Z}_\alpha Z^\alpha+ \nu H^\alpha_\beta{\bar Z}_\alpha
Z^\beta\right)\Big] {\rm d}^{n+1}{\bar Z}\; {\rm d}^{n+1}Z.
\label{eq:15}
\end{eqnarray}

We now observe that we can diagonalise the Hamiltonian by unitary
transformation without affecting any of the terms in (\ref{eq:15})
on account of the fact that every `ket' vector $Z^\alpha$ is
coupled to a corresponding `bra' vector ${\bar Z}_\alpha$.
Therefore, the density of states can be written in the form
\begin{eqnarray}
\fl\Omega(E)=\frac{1}{\pi}\int^{\infty}_{-\infty} \frac{\rd\nu}
{2\pi} \int^{\infty}_{-\infty}\,\frac{{\rm d}\lambda}{2\pi}
\re^{\ri (\lambda+\nu E)} \int_{{\mathbb C}^{n+1}} {\rm
d}^{n+1}\bard{Z}\; {\rm d}^{n+1}Z \exp \left(-{\rm i}\sum_{l=0}^n
(\lambda+\nu E_l)\bard{Z}_l Z^l \right), \label{eq:16}
\end{eqnarray}
where $\{E_l\}_{l=0.1.\ldots,n}$ are the energy eigenstates.

This is of course a formal expression; the integration can
nevertheless be carried out if we regard (\ref{eq:16}) as the
limit of a similar integral in which $\lambda$ and $\nu$ are
displaced into the complex along the negative imaginary axis. The
integration over ${\mathbb C}^{n+1}$ then reduces to a
$(2n-2)$-dimensional Gaussian integral, which is readily performed
to yield
\begin{eqnarray}
\Omega(E)=(-\ri)^{n+1}\pi^n \int_{-\infty}^\infty
\frac{\rd\nu}{2\pi} \int_{-\infty}^\infty
\frac{{\rd}\lambda}{2\pi}\, \re^{\ri(\lambda+\nu E)}
\prod_{l=0}^{n}\frac{1} {(\lambda+\nu E_l)}. \label{eq:17}
\end{eqnarray}
This is the desired integral representation for the density of
states, expressed in terms of the energy eigenvalues.

\section{Density of states for a nondegenerate energy spectrum}
\label{sec:5}

We proceed further by evaluating the integration in (\ref{eq:17})
in the case where the Hamiltonian has no degenerate eigenvalues.
Let us consider the integration in the $\lambda$ variable first.
We observe that there are $n+1$ first order poles on the real
$\lambda$-axis. An application of Cauchy's theorem thus gives us
\begin{eqnarray}
\frac{1}{2\pi\ri} \int^{\infty}_{-\infty}{\rm e}^{\ri(\lambda+\nu
E)}\prod_{l=0}^{n}\frac{1}{(\lambda+\nu E_l)}\rd\lambda =
\sum_{k=0}^n\, {\re}^{\ri\nu(E-E_k)} \prod_{l=0,\neq
k}^{n}\frac{1}{\nu(E_l-E_k)}, \label{eq:18}
\end{eqnarray}
from which it follows that
\begin{eqnarray}
\Omega(E)=\pi^{n}\sum_{k=0}^{n} \int_{-\infty}^{\infty}
\frac{\rd\nu}{2\pi} \frac{\re^{-\ri\nu(E_k-E)}}{(\ri\nu)^n}
\prod_{l=0,\neq k}^{n} \frac{1}{(E_l-E_k)}. \label{eq:19}
\end{eqnarray}
We now recognise the $\nu$-integration formally as the $n$-fold
repeated integral of the $\delta$-function, defined by the
truncated polynomial
\begin{eqnarray}
\delta^{(-n)}(x)= \left\{ \begin{array}{ll}
0 & (x<0) \\
\frac{1}{(n-1)!}\, x^{n-1} & (x\geq 0) .
\end{array} \right.  \label{eq:20}
\end{eqnarray}
As a consequence, the density of states associated with a quantum
system having a nondegenerate energy spectrum can be seen to be
given by an expression of the form given by
\begin{eqnarray}
\Omega(E)=(-1)^n  \pi^{n}\sum_{k=0}^{n}\delta^{(-n)} (E_k-E)
\prod_{l=0,\neq k}^{n}\frac{1}{E_l-E_k}. \label{eq:21}
\end{eqnarray}

In addition to the function $\Omega(E)$ we find it useful for some
purposes to introduce the related normalised density of states
$\mu(E)$ defined by
\begin{eqnarray}
\mu(E)=\frac{\int_\Gamma\delta(H(x)-E)\rd V}{\int_\Gamma\rd V}.
\end{eqnarray}
The function $\mu(E)$ thus satisfies the property that
\begin{eqnarray}
\int_{-\infty}^{\infty} \mu(E)\rd E = 1.
\end{eqnarray}
To calculate the normalisation factor we need to determine the
total volume $V_\Gamma$ of the quantum phase space. This is given
by
\begin{eqnarray}
V_\Gamma = \frac{1}{\pi}\int_{{\mathbb C}^{n+1}} \delta(Z^\alpha
{\bar Z}_\alpha-1) \rd^{n+1}Z \rd^{n+1}{\bar Z},
\end{eqnarray}
where again the factor of $\pi$ in the denominator refers to the
removal of the superfluous overall phase of the wave function.
Thus, making use of the integral representation of the
delta-function, we have
\begin{eqnarray}
V_\Gamma &=& \frac{1}{\pi} \int_{{\mathbb C}^{n+1}} \rd^{n+1}Z
\rd^{n+1}{\bar Z} \int_{-\infty}^\infty \frac{1}{2\pi} \re^{\ri
\lambda(Z^\alpha {\bar Z}_\alpha-1)}\rd\lambda \nonumber \\ &=&
\frac{1}{\pi} \int_{-\infty}^\infty \frac{\rd\lambda}{2\pi}\,
\re^{-\ri\lambda} \int_{{\mathbb C}^{n+1}} \re^{\ri \lambda
Z^\alpha {\bar Z}_\alpha} \rd^{n+1}Z \rd^{n+1}{\bar Z} \nonumber
\\ &=& \frac{1}{\pi} \int_{-\infty}^\infty\frac{\rd\lambda}{2\pi}
\,\re^{-\ri\lambda} \left(-\frac{\ri\pi}{\lambda}\right)^{n+1}
\nonumber \\ &=& \frac{\pi^n}{n!}, \label{eq:23}
\end{eqnarray}
where in the last step we have used the identity
\begin{eqnarray}
\frac{1}{2\pi} \int_{-\infty}^{\infty} \frac{\re^{-\ri\lambda}}
{\lambda^n}\,\rd \lambda = \frac{\ri^n}{(n-1)!}.
\end{eqnarray}
This is in agreement, e.g., with the result obtained in
Ref.~\cite{gibbons} by use of different methods. In particular, we
note that in the case $n=1$ (a two-dimensional Hilbert space) the
space of pure states $\Gamma$ is isomorphic to the geometry of a
round two-sphere with radius one-half.

\section{Equally spaced energy spectrum}
\label{sec:6}

\begin{figure}[th]
\begin{tabular}{ccc}
{\psfig{file=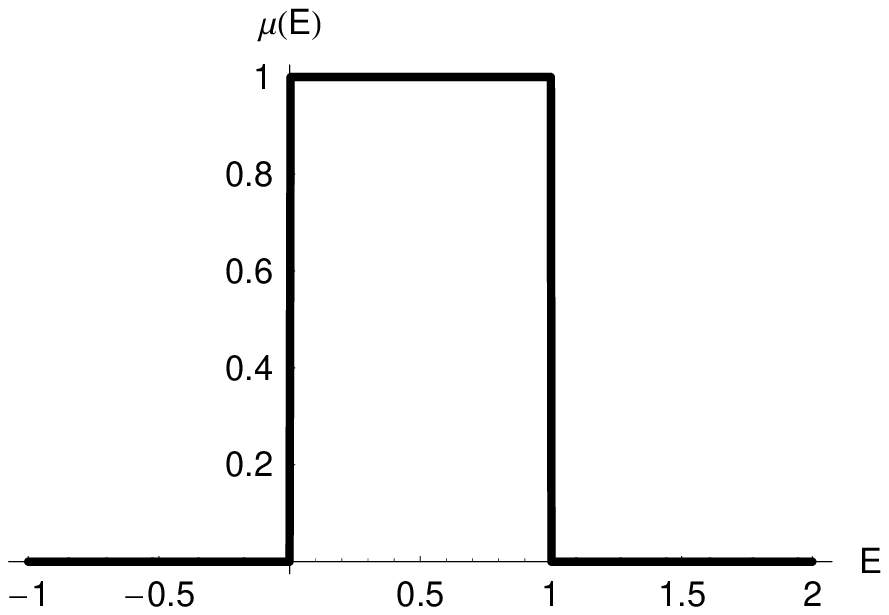,width=5.0cm,angle=0}} \label{fs1} &
{\psfig{file=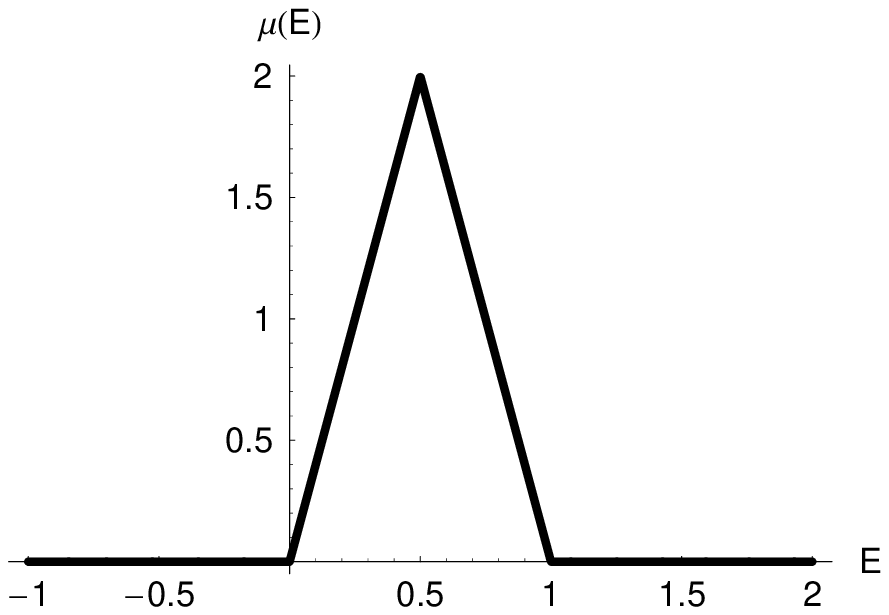,width=5.0cm,angle=0}} \label{fs2} &
{\psfig{file=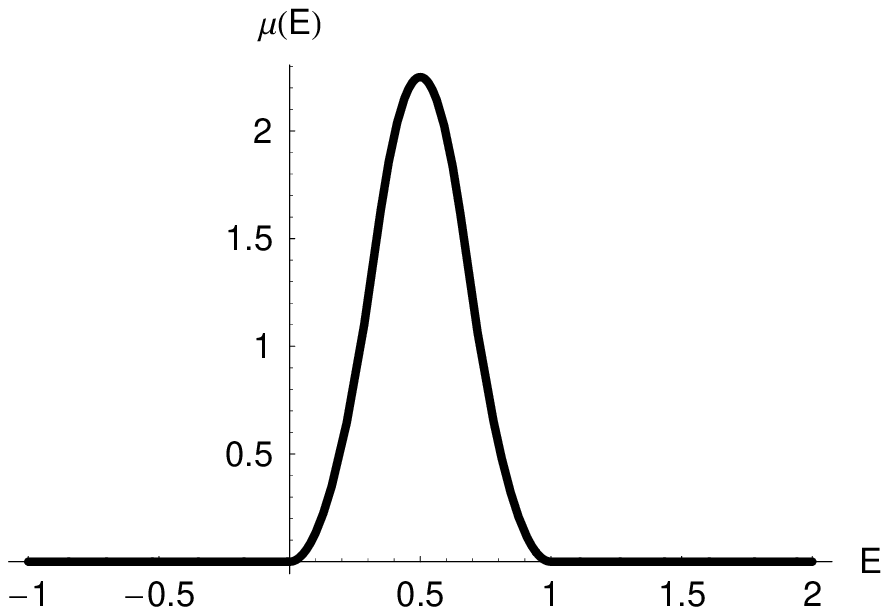,width=5.0cm,angle=0}} \label{fs3}
\end{tabular}
 \caption{
  Density of states as a function of energy $E$ for systems with
  two, three, and four nondegenerate energy eigenstates. The
  functions are evaluated piecewise. For example, in the case of
  the four-dimensional system with
  $E_n/\hbar\omega=0,1,2,3$, the normalised density of states
  $\mu(E)$ is zero for $E\leq0$ and $E> 3$, and is given by three
  distinct quadratic functions in the intervals $(0,1]$, $(1,2]$,
  and $(2,3]$. In general, for an $n$-dimensional system, $\mu(E)$
  is given by a combination of polynomials of degree $n-1$, and
  is at least $n-3$ times differentiable for all
  values of $E$. In all cases the area under the function
  integrates to unity. \label{fig:1}}
\end{figure}

We now specialise the analysis further by considering a quantum
system whose energy eigenvalues are equally spaced. There are many
elementary systems, such as noninteracting particles with spin,
having a spectrum of this kind. To fix the units of energy we set
\begin{eqnarray}
E_k=\hbar\omega k, \qquad (k=0,1,\ldots,n). \label{eq:24}
\end{eqnarray}
Substituting this in (\ref{eq:21}) we obtain
\begin{eqnarray}
\Omega(E)=\frac{(-1)^n\pi^{n}}{(n-1)!} (\hbar\omega)^{n-1} \sum_{k
\geq [E]}^{n}\left(k-\frac{E}{\hbar\omega} \right)^{n-1}
\prod_{l=0,\neq k}^{n}\frac{1}{\hbar\omega(l-k)}. \label{eq:25}
\end{eqnarray}
This expression can be simplified further by writing
\begin{eqnarray}
\prod_{l=0,\neq k}^{n}\frac{1}{\hbar\omega(l-k)}=\frac{1}{(\hbar
\omega)^n (-1)^k k!(n-k)!}, \label{eq:26}
\end{eqnarray}
and hence the microcanonical density of states associated with a
system having equally spaced energy spectrum is given by
\begin{eqnarray}
\Omega(E) = \frac{(-1)^n}{(n-1)!} \pi^{n}
\frac{1}{\hbar\omega}\sum_{k \geq [E]}^{n} \frac{(-1)^k
\left(k-E/\hbar\omega\right)^{n-1}}{ k!(n-k)!}. \label{eq:27}
\end{eqnarray}
Finally, dividing this expression by the normalisation factor
(\ref{eq:23}) we obtain the following formula for the normalised
density of states:
\begin{eqnarray}
\mu(E) = \frac{(-1)^n n}{\hbar\omega} \sum_{k \geq [E]}^{n}
\frac{(-1)^k \left(k-E/\hbar\omega\right)^{n-1}}{ k!(n-k)!}.
\label{eq:28}
\end{eqnarray}
It will be useful now to indicate what this function looks like in
various situations. In Figure~\ref{fig:1} we plot $\mu(E)$ for
two, three, and four dimensional Hilbert spaces.

\begin{figure}[th]
{\centerline{\psfig{file=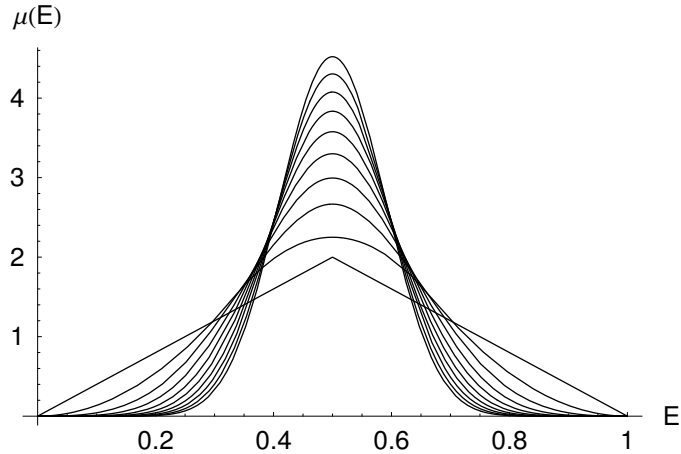,width=10cm,angle=0}}}
 \caption{Rescaled normalised density of states $\mu(E)$ as a
 function of energy $E$. As we increase the density of energy
 levels in the interval $[0,1]$, the corresponding density of
 states becomes more sharply distributed around the centre
 $E=\frac{1}{2}$. For these numerical plots we have set
 $\hbar\omega=1$. The plots correspond to systems for which the
 number of energy levels ranges from $3$ to $12$, and the value of
 the rescaled energy varies over the unit interval.
 \label{fig:2}}
\end{figure}

We would like now to study the behaviour of the density of states
as we increase the number of energy levels. For this purpose we
find it convenient to rescale the energy spectrum so that the
range of energy is over a fixed interval $[0,1]$. After the
application of a suitable such rescaling, the density of states
\label{eq:28} reduces to
\begin{eqnarray}
\mu(E) = \frac{(-1)^n n^2}{\hbar\omega} \sum_{k \geq [E]}^{n}
\frac{(-1)^k \left(k- nE/\hbar\omega \right)^{n-1}}{ k!(n-k)!}.
\label{eq:29}
\end{eqnarray}
In Figure~\ref{fig:2} we plot this rescaled density of states for
a variety of systems with different numbers of energy levels.

\begin{figure}[th]
{\centerline{\psfig{file=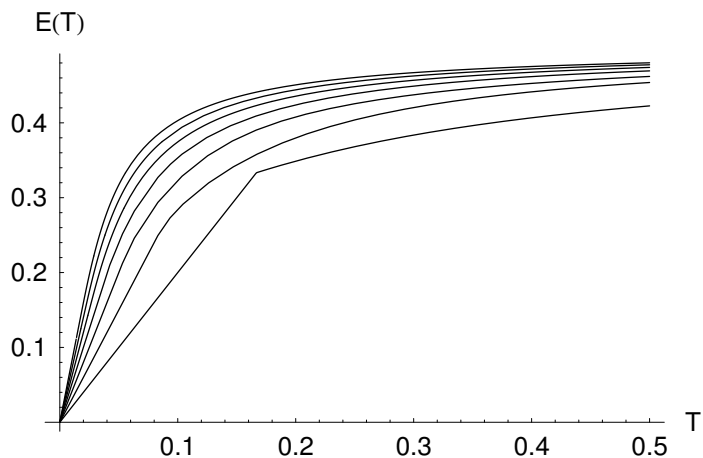,width=10cm,angle=0}}}
 \caption{Energy $E(T)$ as a function of temperature $T$. The
 linear curve with lowest energy corresponds to a four-level
 system,
 and we have plotted $E(T)$ up to the ten-level system with an
 interval of 1. Although not shown in the plot, the second root
 for $E(0)$ is given by $E(0)=1$, where the function is
 multi-valued. As $T$ is reduced from zero, $E(T)$ reduces
 monotonically, and approaches $E=\frac{1}{2}$ as $T\to-\infty$.
 This is the region inaccessible in equilibrium. \label{fig:3}}
\end{figure}

Expression (\ref{eq:29}) for the density of states also allows us
to study the relation between energy $E$ and temperature $T$, by
use of the relation (\ref{eq:6}). In Figure~\ref{fig:3} we plot
the system energy $E(T)$ as a function of temperature by
numerically inverting the relation
\begin{eqnarray}
T = \left(\frac{\rd\Omega}{\rd E}\right)^{-1}\Omega(E).
\label{eq:30}
\end{eqnarray}
As the temperature increases from zero, the corresponding energy
increases monotonically, and asymptotically approaches the value
$E= \frac{1}{2}$ (in the rescaled units). The function $E(T)$ is
multi-valued at $T=0$, where the energy takes the values $E=0,1$.
In the region $E\in(\frac{1}{2},1]$ the density of states $\mu(E)$
is a decreasing function of $E$, and thus the corresponding
`temperature' is negative. To put the matter differently, this is
the region that is not accessible in an equilibrium.

\section{Convergence in the infinite energy-level limit}
\label{sec:7}

In statistical mechanics it is often presumed that the various
different distributions used to describe equilibrium states (e.g.,
microcanonical, canonical, grandcanonical, and
pressure-temperature distributions) should in some respects be
equivalent in the infinite volume or thermodynamic limit. Now the
system that we have studied in the previous section corresponds to
a system of noninteracting quantum particles. As a consequence, we
would not expect any nontrivial behaviour exhibited by the system
in the thermodynamic limit. Nevertheless it would be of interest
to study how the density of states behaves in this limit so that
the result might be compared with the corresponding canonical
formulation outlined in Ref.~\cite{brody2}.

Now the results shown in Figure~\ref{fig:2} indicate that as we
increase the number of energy levels the density of states becomes
more and more peaked at the intermediate energy $E= \frac{1}{2}$.
Therefore, in this section we consider the separation of a pair of
normalised state densities associated with $k$-level and
$(k+1)$-level systems for a range of values for $k$, and study
whether the separation decreases as we increase $k$. If so, then
the result will indicate that the density function is converging,
possibly to a delta-function centred at $E= \frac{1}{2}$.

There are various standard measures that one can use to study the
separation of a pair of density functions, such as the relative
entropy or the Bhattacharyya distance. Here we shall consider a
closely related measure given by
\begin{eqnarray}
D(\mu_m,\mu_n) = \sqrt{2-2\int_0^1 \sqrt{\mu_n(E)} \sqrt{\mu_m(E)}
\rd E}, \label{eq:31}
\end{eqnarray}
known as the Hellinger distance. For clarity we let $\mu_k(E)$
denote the density of states associated with a $k$-level system.
Note that the Hellinger distance is simply the $L^2$-norm
\begin{eqnarray}
D(\mu_m,\mu_n) =\left\|\sqrt{\mu_n(E)}-\sqrt{\mu_m(E)}\right\|
\label{eq:32}
\end{eqnarray}
of the difference between the two square-root density functions.

\begin{figure}[th]
{\centerline{\psfig{file=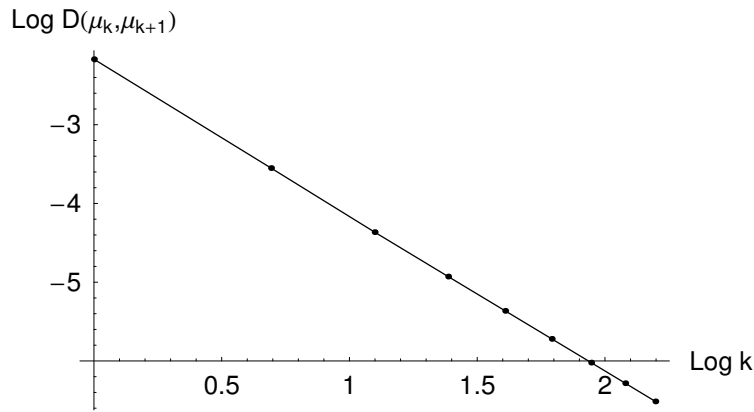,width=10cm,angle=0}}}
 \caption{The Hellinger distance $D(\mu_k, \mu_{k+1})$ between
 two `consecutive' rescaled state densities is plotted for
 $k=2,3,\ldots,10$. The plot is presented on a logarithmic scale.
 We find that the logarithms of the distances lie almost perfectly
 on a straight line with gradient of $-2$, indicating that the
 separation of the densities decays quadratically with an
 increasing number of energy levels. \label{fig:4}}
\end{figure}

The result of the numerical analysis of (\ref{eq:31}) is shown in
Figure~\ref{fig:4}, which indicates that the relative separation
of $\mu_k(E)$ and $\mu_{k+1}(E)$ is decreasing quadratically in
$k$. Therefore, we conclude that the density of states is likely
to be converging in the sense of $L^2$.

\section{Other spectral structures}
\label{sec:8}

\begin{figure}[th]
\begin{tabular}{ccc}
{\psfig{file=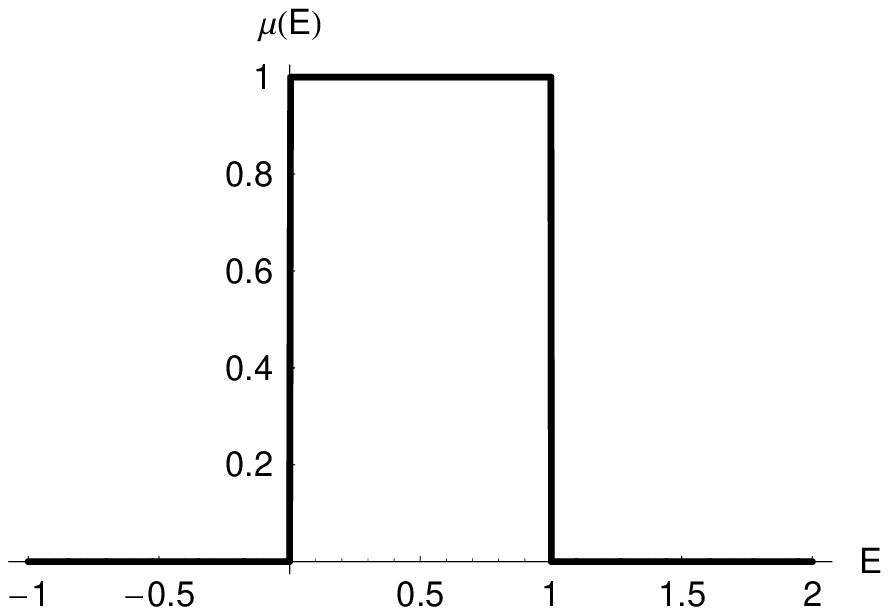,width=5.0cm,angle=0}} \label{qs1} &
{\psfig{file=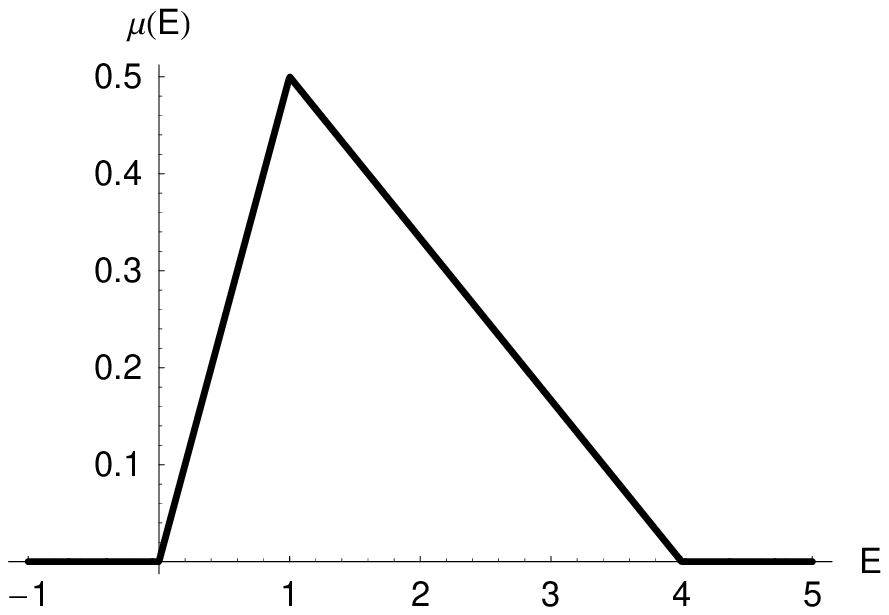,width=5.0cm,angle=0}} \label{qs2} &
{\psfig{file=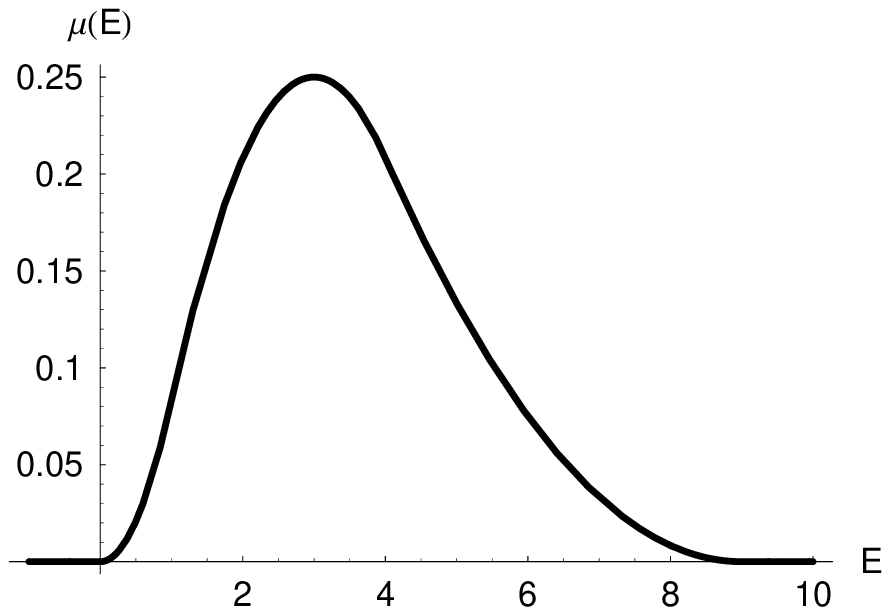,width=5.0cm,angle=0}} \label{qs3}
\end{tabular}
 \caption{The density of states $\mu_k(E)$ for $k$-level systems
 with $k=2,3,4$, in the case of a system with a quadratically
 growing energy spectrum. For the plot we have set
 $\hbar\omega=1$. Unlike the example with a linear, equally-spaced
 spectrum, the density of states is not symmetric. \label{fig:5}}
\end{figure}

In the example considered in \S\ref{sec:6} and \S\ref{sec:7} above
we analysed a linear spectral structure of the form $E_k\propto
k$, and found that the density of states is centred at an
intermediate energy value. In this section we investigate
nondegenerate systems having other kinds of energy growths. We
start with the case of a system whose energy grows quadratically
so that $E_k=\hbar\omega k^2$. In this case we make use of the
relation
\begin{eqnarray}
\prod_{l=0,\neq k}^{n} \frac{1}{l^2-k^2}=\prod_{l=0,\neq k}^{n}
\frac{1}{(l+k)(1-k)}=\frac{2(-1)^k}{(n+k)!(n-k)!}, \label{eq:39}
\end{eqnarray}
and substitute this in formula (\ref{eq:21}) to obtain
\begin{eqnarray}
\mu(E) = \frac{2n(-1)^n}{\hbar\omega} \sum_{k\geq[E]}^{n}
\frac{(-1)^k \left(k^2-\frac{E}{\hbar\omega}\right)^{n-1}}{
(n+k)!(n-k)!}. \label{eq:40}
\end{eqnarray}
As one might have expected, the functions given by (\ref{eq:40})
have similar characteristics to those of the previous example. For
each value of $n$ we obtain a curve $\mu(E)$ which is specified by
$n+2$ polynominals of degree $n-1$. The function is continuous and
its differentiability is of order $n-2$. The first three examples
are plotted in Figure~\ref{fig:5}. As the plots indicate, the
microcanonical density functions in the current examples are no
longer symmetric around the intermediate energy
$\frac{1}{2}(E_{\rm min}+E_{\rm max})$, in contrast to the case of
linear energy growth.

\begin{figure}[th]
\centerline{\psfig{file=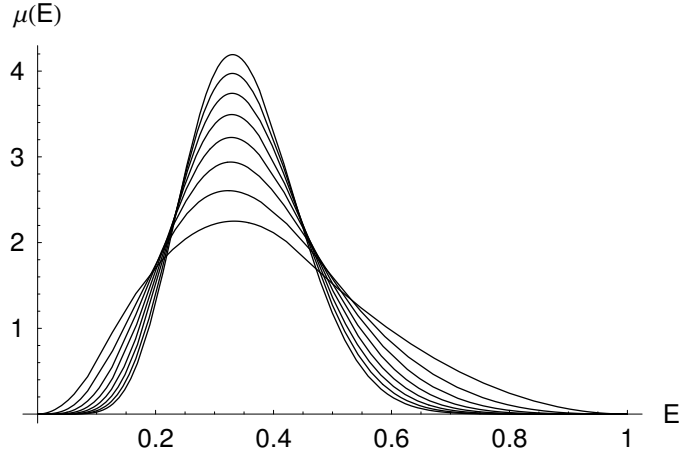,width=10.0cm,angle=0}}
 \caption{The density of states $\mu(E)$ associated with a system
 having the spectrum $E_k=k^2/(N-1)^2$, where $N$ denotes
 total number of energy levels. The value of $k$ thus ranges over
 $k=0,1,\ldots,N-1$. Here we plot $\mu(E)$ for $N=4,5,
 \ldots,11$. Numerical studies show that the peaks of the
 distributions are located at approximately $E=\frac{1}{3}$.
 \label{fig:6}}
\end{figure}

As in the previous example, in the present case we can rescale the
energy spectrum in such a way that we can directly compare the
behaviour of the density of states as we increase the number of
energy levels. The result is shown in Figure~\ref{fig:6}. We
observe that the density of states becomes more peaked as we
increase the number of energy levels. However, the location of the
peak is no longer at $E=\frac{1}{2}$, but rather closer to
$E=\frac{1}{3}$.

\begin{figure}[th]
\centerline{\psfig{file=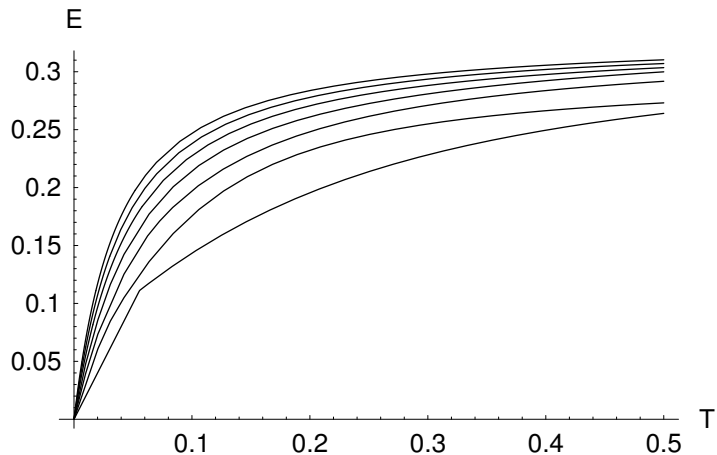,width=10.0cm,angle=0}}
 \caption{System energy $E(T)$ as a function of temperature $T$
 in the case of a system having a quadratic energy spectrum. The
 plots correspond to a set of $k$-level systems with
 $k=5,6,\ldots,11$.
 Although not shown in the plot, the energy is multi-valued at
 $T=0$ so that $E(T)\to0$ as $T\to0^+$ and $E(T)\to1$ as $T\to
 0^-$. The range of energy accessible in equilibrium is thus given
 by $0\leq E\lesssim\frac{1}{3}$.  \label{fig:7}}
\end{figure}

As a consequence of the skewed form of the distribution, the range
of energy for which the derivative of $\mu(E)$ with respect to $E$
remains positive is somewhat reduced. This implies that the range
of energy associated with positive temperature is reduced from the
previous example of linear energy spectrum. Some examples are
illustrated in Figure~\ref{fig:7} where we plot the temperature
dependence of the energy. As the temperature is increased, the
energy grows monotonically and reaches a value around
$\frac{1}{3}$. The remaining values of the energy are associated
with negative temperatures that are inaccessible in equilibrium.

We have examined systems having a linear energy growth and a
quadratic energy growth. In the linear case the microcanonical
distribution is symmetric around its centre, while in the
quadratic case the peaks of the distributions have shifted to the
left with smaller energies. This leads to the question of what
happens to the density of states associated with systems having
other spectral structures.

\begin{figure}[th]
\centerline{\psfig{file=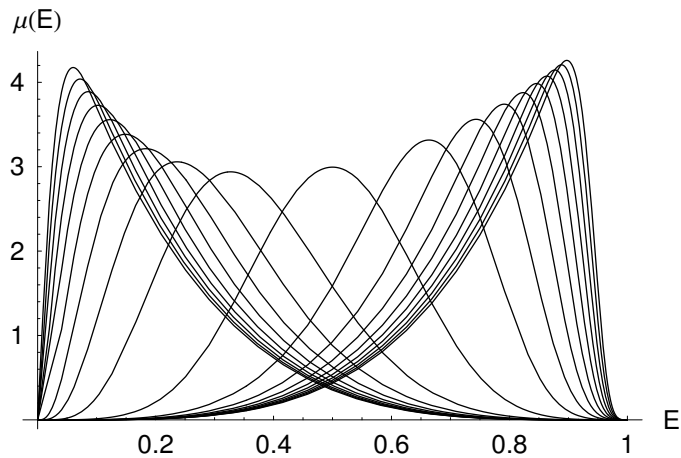,width=10.0cm,angle=0}}
 \caption{The density of states $\mu(E)$ associated with systems
 having the spectra $E_n=n^k$ and $E_n=
 n^{\frac{1}{k}}$, for $k=1,2,\ldots,10$. In the former case the
 peak of $\mu(E)$ shifts to the left, while in the latter case the
 peak shifts to the right. The cases $E_n=n^k$ and $E_n=
 n^{\frac{1}{k}}$ are close to but not exactly symmetric to one
 another around $E=\frac{1}{2}$. In all examples the number of
 energy levels is set to six. \label{fig:8}}
\end{figure}

In the case of a system with an infinite number of degrees of
freedom (and thus an infinite number of energy levels), the energy
spectrum cannot grow more rapidly than quadratically in the number
of energy levels. However, for a finite system there is in
principle no limitation on how fast the system energy can grow.
Therefore, we would like to study the behaviour of the density of
states associated with systems having the following two spectral
structures:
\begin{eqnarray}
E_n = n^k \qquad {\rm and} \qquad E_n = n^{1/k}, \label{eq:41}
\end{eqnarray}
for $k=1,2,3,\ldots$. Thus in one case the growth of the energy is
enhanced as $k$ is increased, while in the other case the growth
of the energy is suppressed. Various densities of states
associated with systems having spectra of the form (\ref{eq:41})
are computed numerically and plotted in Figure~\ref{fig:8}. The
result shows that as the rate of growth is increased, the location
of the peak of $\mu(E)$ becomes smaller, whereas when the rate of
energy growth is suppressed, the peak of $\mu(E)$ increases. In
particular, the peak of $\mu(E)$ for the system with energy
$E_n=n^k$ is located approximately around $E\sim(k+1)^{-1}$, while
the peak of $\mu(E)$ for the system with energy $E_n=n^{1/k}$ is
located approximately around $E\sim1-(k+1)^{-1}$. Therefore, when
$k\ll1$, the values of energy that are accessible in equilibrium
becomes negligible when $E_n=n^k$, whereas if
$E_n=n^{\frac{1}{k}}$ then virtually all values of $E\in[0,1]$ are
accessible.

In conclusion, we note that the analysis we have pursued here is
based on a generalised microcanonical postulate, which relaxes the
somewhat more stringent assumptions made in the standard quantum
microcanonical postulate as outlined in, e.g., Ref.~\cite{huang}.
To determine whether the assumption made in this paper reflects
the actual equilibrium distribution of an isolated quantum system
it will be desirable to examine the properties of interacting
systems. In the case of a general interacting system, some of the
energy levels are typically highly degenerate, and thus we must
return to the integral representation (\ref{eq:17}) with a view to
deriving an efficient way to carry out the integration in the
cases when there are higher-order poles. This is an intriguing
open problem that we hope to investigate elsewhere.

\vspace{0.5cm}
\begin{footnotesize}
\noindent DCB acknowledges support from The Royal Society.
\end{footnotesize}

\end{document}